\documentclass[twocolumn]{revtex4-1}

\usepackage{amsmath}
\usepackage[colorlinks,citecolor=blue,linkcolor=blue]{hyperref}
\usepackage{multirow}
\usepackage{graphicx}
\usepackage{epstopdf}

\begin{document}

\title{
Determination of $f_+^K(0)$ and Extraction of
$|V_{cs}|$ from Semileptonic $D$ Decays}
\author{Y. Fang}
\author{G. Rong}
\author{H. L. Ma}
\author{J. Y. Zhao}
\affiliation{Institute of High Energy Physics, Beijing 100039, People's Republic of China}
\date{\today}

\begin{abstract}
By globally analyzing all existing measured branching fractions and partial rates in different four momentum transfer-squared $q^2$ bins of
$D\to Ke^+\nu_e$ decays, we obtain the product of the form factor and magnitude of
CKM matrix element $V_{cs}$ to be $f_+^K(0)|V_{cs}|=0.717\pm0.004$.
With this product, we determine the $D\to K$ semileptonic form factor $f_+^K(0)=0.737\pm0.004\pm0.000$ in conjunction with the value of $|V_{cs}|$ determined from the SM global fit.
Alternately, with the product together with the input of the form factor $f_+^K(0)$  calculated in lattice QCD recently,
we extract $|V_{cs}|^{D\to Ke^+\nu_e}=0.962\pm0.005\pm0.014$, where the error is still dominated by
the uncertainty of the form factor calculated in lattice QCD.
Combining the $|V_{cs}|^{D_s^+\to\ell^+\nu_\ell}=1.012\pm0.015\pm0.009$ extracted from all existing measurements of $D^+_s\to\ell^+\nu_\ell$ decays and $|V_{cs}|^{D\to Ke^+\nu_e}=0.962\pm0.005\pm0.014$ together, we find the most precisely determined $|V_{cs}|$ to be $|V_{cs}|=0.983\pm0.011$,
which improves the accuracy of the PDG'2014 value $|V_{cs}|^{\rm PDG'2014}=0.986\pm0.016$ by $45\%$.
\end{abstract}

\maketitle

\section{Introduction}
\label{sec:intro}

In the Standard Model (SM) of particle physics, the mixing between the quark flavours in weak interaction is parameterized by the Cabibbo-Kobayashi-Maskawa (CKM) matrix, which
is a $3\times3$ unitary matrix.
Since the CKM matrix elements are fundamental parameters of the SM, precise determinations of these elements are necessary and very important in testing the SM and searching for New Physics (NP).

Since the effects of strong interactions and weak interaction can be well separated in
semileptonic $D$ decays, these decays are excellent processes from which we can determine the magnitude of CKM matrix element $V_{cs(d)}$.
In the SM, neglecting the lepton mass, the tree-level
differential decay rate in absence of radiative correction
for $D\to Ke^+\nu_e$ process is given by
\begin{equation}\label{eq:dG_dq2}
    \frac{d\Gamma}{dq^2} = \frac{G_F^2}{24\pi^3}|V_{cs}|^2 \boldsymbol{p}^3|f_+^K(q^2)|^2,
\end{equation}
where
$G_F$ is the Fermi constant, $\boldsymbol p$ is the three momentum of the $K$ meson in the rest frame of the $D$ meson, $q^2$ is the four momentum transfer-squared, i.e. the invariant mass of the lepton and neutrino system,
and $f_+^K(q^2)$ is the form factor which parameterizes the effect of strong interaction.

In addition to extraction of $|V_{cs}|$,
the precise measurements of the $D\to K$ semileptonic form factor is also very important to validate the lattice QCD (LQCD) calculation of the form factor.
If the LQCD calculation of the form factor pass the test with the precisely measured form factor for $D\to Ke^+\nu_e$ decay, the uncertainty of the semileptonic $B$ decay form factor calculated in LQCD would be reduced.
This would help in reducing the uncertainty of the measured $|V_{ub}|$ from semileptonic $B$ decays~\cite{RongG_program_ccbar}.
The improved measurement of $|V_{ub}|$ from semileptonic $B$ decay will improve the determination of the $B_d$ unitarity triangle, with which one can more precisely test the SM and search for NP.

In the past decades, copious measurements of branching fractions and/or decay rates for $D\to Ke^+\nu_e$ decays were performed at more than ten experiments.
By comprehensive analysis of these existing measurements together with $|V_{cs}|$ from SM global fit or
together with form factor $f_+^K(0)$ calculated in LQCD, one can precisely
determine the form factor $f_+^K(0)$ or extract $|V_{cs}|$.

In this article, we report the determination of $f_+^K(0)$ or extraction of $|V_{cs}|$ by analyzing all of these existing measurements of the
semileptonic $D\to Ke^+\nu_e$ decays in conjunction with $|V_{cs}|$ from SM global fit or with the form factor $f_+^K(0)$
calculated in lattice QCD.
In the following sections, we first review the experimental measurements of branching fractions and decay rates for $D\to Ke^+\nu_e$ decays in Section~\ref{sec:expt}.
We then describe our comprehensive analysis procedure for dealing with these measurements to obtain the product of $f_+^K(0)$ and $|V_{cs}|$ in Section~\ref{sec:ana}.
In Section~\ref{sec:rslt}, we present the final results of our comprehensive analysis of these measurements.
We finally give a summary for the determination of $f_+^K(0)$ and the extraction of $|V_{cs}|$ in Section~\ref{sec:sum}.

\section{Experiments}
\label{sec:expt}

\subsection{Relative Measurements}

In 1989, the Tagged Photon Spectrometer Collaboration studied the $D^0\to K^-e^+\nu_e$ decays and found 250 signal events for $D^0\to K^-e^+\nu_e$ decays at the E691 experiment.
Based on these events, they measured the ratio of decay rates
$R_0 \equiv \Gamma(D^0\to K^-e^+\nu_e)/\Gamma(D^0\to K^-\pi^+)=0.91\pm0.07\pm0.11$~
\cite{E691}.

In 1991, by analyzing 490 pb$^{-1}$ data collected with the CLEO detector at the Cornell Electron Storage Ring (CESR), the CLEO Collaboration made a measurement of the branching ratio of $D^0$ semileptonic decays.
They observed $584\pm37\pm39$ signal events from $D^0\to K^-e^+\nu_e$ decays and
obtained the ratio of branching fractions $R_0 \equiv B(D^0\to K^-e^+\nu_e)$ $/B(D^0\to K^-\pi^+)$ $=0.90\pm0.06\pm0.06$~\cite{CLEO}.

In 1993, the CLEO Collaboration measured the branching ratios of the semileptonic $D$
decay modes using $1.68$ fb$^{-1}$ data collected with the CLEO-II detector at
CESR.
They selected the semileptonic $D$ decays from $e^+e^-\to c\bar c$ events and measured the ratios $R_0 \equiv B(D^0\to K^-e^+\nu_e)$ $/B(D^0\to K^-\pi^+)$ $=0.978\pm0.027\pm0.044$ and $R_+ \equiv B(D^+\to\bar K^0e^+\nu_e)$ $/B(D^+\to\bar K^0\pi^+)$
$=2.60\pm0.35\pm0.26$ \cite{CLEOII}.

In 2007, the BaBar Collaboration studied the $D^0\to K^-e^+\nu_e$ decays by analyzing
75 fb$^{-1}$ data collected at 10.6 GeV~\cite{BaBar}.
They selected $D^0\to K^-e^+\nu_e$ decays
from $e^+e^-\to c\bar c$ events and divide the candidate events into ten $q^2$ bins.
In each $q^2$ bin, the partial decay rate is measured
relative to the normalization mode, $D^0\to K^-\pi^+$.

All above mentioned measurements are relative measurements which could not be used directly to determine the form factor $f_+^K(0)$ or $|V_{cs}|$. To use these measurements to determine $f_+^K(0)$ or $|V_{cs}|$, we should first transfer these measurements into absolute decay rates in certain $q^2$ range. The absolute decay rate $\Delta\Gamma$ can be obtained from the measured relative decay branching ratio $R$ by
\begin{equation}
\Delta\Gamma=R\times B(D\to K\pi)\times\frac{1}{\tau_D},
\end{equation}
where
$B(D\to K\pi)$ is the branching fraction for $D^0\to K^-\pi^+$ or $D^+\to \bar K^0\pi^+$ decays,
and $\tau_D$ is the lifetime of $D$ meson.
To avoid the possible correlations, here we use the value of the branching fraction of $D^0\to K^-\pi^+$ decay,
$B(D^0\to K^-\pi^+)=(3.91\pm0.05)\%$,
which is the average of the measurements from
BaBar~\cite{B_D0toKpi_BaBar},
CLEO-c~\cite{B_D0toKpi_CLEOc},
CLEO-II~\cite{B_D0toKpi_CLEOII},
ALEPH~\cite{B_D0toKpi_ALEP,B_D0toKpi_ALEP2},
and
ARGUS~\cite{B_D0toKpi_ARG}.
For the branching fraction of $D^+\to\bar K^0\pi^+$ decay,
we use the value of $B(D^+\to \bar K^0\pi^+)=(2.986\pm0.069)\%$,
which is the sum of CLEO-c's measurements
$B(D^+\to K_S^0\pi^+)=(1.526\pm0.022\pm0.038)\%$~\cite{B_D0toKpi_CLEOc}
and
$B(D^+\to K_L^0\pi^+)=(1.460\pm0.040\pm0.035)\%$~\cite{B_DptoKLpi_CLEOc}.
Using the lifetime of $D$ meson, $\tau_{D^0}=(410.1\pm1.5)\times10^{-15}$ s, and $\tau_{D^+}=(1040\pm7)\times10^{-15}$ s from PDG~\cite{pdg},
the branching fractions of $B(D^0\to K^-\pi^+)=(3.91\pm0.05)\%$ and $B(D^+\to \bar K^0\pi^+)=(2.986\pm0.069)\%$, we translate these measurements of relative branching fractions and relative partial decay rates into absolute partial decay rates as shown in Tabs.~\ref{tab:Expt_D0} and \ref{tab:Expt_Dp}.

%%%%%%%%%%%%%%%%%%%%%%%%%%%%%%%%%%%%%%%%%%%%%%%%%%%%%%%%%%%%%%%%%%%%%%%%
\begin{table}[h]
\caption{The partial rates $\Delta\Gamma$ of the $D^0\to K^-e^+\nu_e$ decays in $q^2$ ranges obtained from different experiments.
$q^2_{\rm max}$ is the maximum value of $q^2$.
}
\label{tab:Expt_D0}
\begin{center}
%\resizebox{0.48\textwidth}{!}{
\begin{tabular}{llr}
\hline\hline
Experiment & $q^2$ (GeV$/c^2$) & $\Delta\Gamma$ (ns$^{-1}$)  \\
\hline
 E691~\cite{E691} & (0.0, $q^2_{\rm max}$) & $86.76\pm12.48$ \\
 CLEO~\cite{CLEO} & (0.0, $q^2_{\rm max}$) & $85.81\pm 8.17$ \\
 CLEO-II~\cite{CLEOII} & (0.0, $q^2_{\rm max}$) & $93.25\pm 5.08$ \\
\hline
 BaBar~\cite{BaBar}
 & (0.0, 0.2) & $17.75\pm0.48$ \\
 & (0.2, 0.4) & $16.26\pm0.49$ \\
 & (0.4, 0.6) & $14.42\pm0.42$ \\
 & (0.6, 0.8) & $12.39\pm0.38$ \\
 & (0.8, 1.0) & $ 9.92\pm0.31$ \\
 & (1.0, 1.2) & $ 7.72\pm0.26$ \\
 & (1.2, 1.4) & $ 5.32\pm0.21$ \\
 & (1.4, 1.6) & $ 3.24\pm0.14$ \\
 & (1.6, 1.8) & $ 1.29\pm0.09$ \\
 & (1.8, $q^2_{\rm max}$) & $ 0.06\pm0.01$ \\
\hline
 Mark-III~\cite{MarkIII} & (0.0, $q^2_{\rm max}$) & $82.91\pm15.62$ \\
 BES-II~\cite{BESII_D0} & (0.0, $q^2_{\rm max}$) & $93.15\pm11.77$ \\
 BES-III~\cite{BESIII_D0Kenu} & (0.0, $q^2_{\rm max}$) & $85.47\pm0.93$ \\
\hline
 CLEO-c~\cite{CLEOc}
 & (0.0, 0.2) & $17.82\pm0.43$ \\
 & (0.2, 0.4) & $15.83\pm0.39$ \\
 & (0.4, 0.6) & $13.91\pm0.36$ \\
 & (0.6, 0.8) & $11.69\pm0.32$ \\
 & (0.8, 1.0) & $ 9.36\pm0.28$ \\
 & (1.0, 1.2) & $ 7.08\pm0.24$ \\
 & (1.2, 1.4) & $ 5.34\pm0.21$ \\
 & (1.4, 1.6) & $ 3.09\pm0.16$ \\
 & (1.6, $q^2_{\rm max}$) & $ 1.28\pm0.11$ \\
\hline\hline
\end{tabular}%}
\end{center}
\end{table}
%%%%%%%%%%%%%%%%%%%%%%%%%%%%%%%%%%%%%%%%%%%%%%%%%%%%%%%%%%%%%%%%%%%%%%%%

\begin{table}[h]
\caption{The partial rates of the $D^+\to \bar K^0 e^+\nu_e$ decays in $q^2$ ranges obtained from different experiments.
$q^2_{\rm max}$ is the maximum value of $q^2$.
}
\label{tab:Expt_Dp}
\begin{center}
%\resizebox{0.48\textwidth}{!}{
\begin{tabular}{llr}
\hline\hline
Experiment & $q^2$ (GeV$/c^2$) & $\Delta\Gamma$ (ns$^{-1}$) \\
\hline
 CLEO-II~\cite{CLEOII} & (0.0, $q^2_{\max}$) & $74.65\pm12.65$ \\
 BES-II~\cite{BESII_Dp} & (0.0, $q^2_{\rm max}$) & $86.06\pm16.60$ \\
\hline
 CLEO-c~\cite{CLEOc}
 & (0.0, 0.2) & $17.79\pm0.65$ \\
 & (0.2, 0.4) & $15.62\pm0.59$ \\
 & (0.4, 0.6) & $14.02\pm0.54$ \\
 & (0.6, 0.8) & $12.28\pm0.49$ \\
 & (0.8, 1.0) & $ 8.92\pm0.41$ \\
 & (1.0, 1.2) & $ 8.17\pm0.37$ \\
 & (1.2, 1.4) & $ 4.96\pm0.27$ \\
 & (1.4, 1.6) & $ 2.67\pm0.19$ \\
 & (1.6, $q^2_{\rm max}$) & $ 1.19\pm0.13$ \\
\hline\hline
\end{tabular}%}
\end{center}
\end{table}
%%%%%%%%%%%%%%%%%%%%%%%%%%%%%%%%%%%%%%%%%%%%%%%%%%%%%%%%%%%%%%%%%%%%%%%%

In addition to the measurements of relative branching fractions and relative partial rates, the FOCUS Collaboration measured the non-parametric relative form factors $f_+^K(q^2)/f_+^K(0)$ at the central values of nine $q^2$ bins by analyzing the $D^0\to K^-\mu^+\nu_\mu$ decays in 2005~\cite{FOCUS}.
These measured variations of $f_+^K(q^2)/f_+^K(0)$ at FOCUS experiment also provide
useful information about the semileptonic decay form factor and are helpful to determine the product $f_+^K(0)|V_{cs}|$ and the shape parameters of the form factor.
These measurements are listed in Tab.~\ref{tab:Expt_ff_FOCUS} and are used in the further analysis.

\begin{table}[h]
\caption{Measurements of normalized form factors $f_+^K(q^2_i)/f_+^K(0)$ at the FOCUS experiment.
}
\label{tab:Expt_ff_FOCUS}
\begin{center}
%\resizebox{0.48\textwidth}{!}{
\begin{tabular}{lcc}
\hline\hline
$i$ & $q^2_i$ (GeV$/c^2$) & $f_+^K(q^2_i)/f_+^K(0)$ \\
\hline
     1 & 0.09 & $1.01 \pm 0.03$ \\
     2 & 0.27 & $1.11 \pm 0.05$ \\
     3 & 0.45 & $1.15 \pm 0.07$ \\
     4 & 0.63 & $1.17 \pm 0.08$ \\
     5 & 0.81 & $1.24 \pm 0.09$ \\
     6 & 0.99 & $1.45 \pm 0.09$ \\
     7 & 1.17 & $1.47 \pm 0.11$ \\
     8 & 1.35 & $1.48 \pm 0.16$ \\
     9 & 1.53 & $1.84 \pm 0.19$ \\
\hline\hline
\end{tabular}%}
\end{center}
\end{table}
%%%%%%%%%%%%%%%%%%%%%%%%%%%%%%%%%%%%%%%%%%%%%%%%%%%%%%%%%%%%%%%%%%%%%%%%

\subsection{Absolute Measurements}

In 1989, the Mark III Collaboration performed a measurement of absolute branching fraction for semileptonic $D^0\to K^-e^+\nu_e$ decay by analyzing data taken at the peak of $\psi(3770)$ resonance with the Mark III detector.
They tagged $3636\pm54\pm195$ $\bar D^0$ mesons and found $55$ $D^0\to K^-e^+\nu_e$ signal events in the system recoiling against the $\bar D^0$ tags.
With these events, they measured the absolute decay branching fraction $B(D^0\to K^-e^+\nu_e)=(3.4\pm0.5\pm0.4)\%$~\cite{MarkIII}.

Using the similar method as the one used in Mark III, the BES-II Collaboration measured the branching fractions of $D\to Ke^+\nu_e$ decays by analyzing about 33 pb$^{-1}$ data taken near 3.773 GeV with the BES-II detector at the BEPC collider. Their results are $B(D^0\to K^-e^+\nu_e)=(3.82\pm0.40\pm0.27)\%$~\cite{BESII_D0} and $B(D^+\to \bar K^0e^+\nu_e)=(8.95\pm1.59\pm0.67)\%$~\cite{BESII_Dp}.

Recently, the BESIII Collaboration reported preliminary results of
$D^0\to K^-e^+\nu_e$ decays obtained by analyzing
2.92 fb$^{-1}$ data taken at 3.773 GeV.
They accumulated $(279.3\pm0.4)\times10^4$ $\bar D^0$ tags from five hadronic decay modes.
In this sample of $\bar D^0$ tags, they observed $70727\pm278$ signal events for $D^0\to K^-e^+\nu_e$ decays and measured the branching fraction
$B(D^0\to K^-e^+\nu_e)=(3.505\pm0.014\pm0.033)\%$~\cite{BESIII_D0Kenu}.

The partial decay rate is related to the decay branching fraction by
\begin{equation}
\Delta\Gamma=B(D\to Ke^+\nu_e)\times\frac{1}{\tau_D}.
\end{equation}
Using the lifetimes of $D^0$ and $D^+$ mesons quoted from PDG~\cite{pdg}, $\tau_{D^0}=(410.1\pm1.5)\times10^{-15}$ s and $\tau_{D^+}=(1040\pm7)\times10^{-15}$ s, we translate these absolute measurements of branching fractions for $D\to Ke^+\nu_e$ decays into the partial decay rates, which are shown in Tabs.~\ref{tab:Expt_D0} and \ref{tab:Expt_Dp}.

In 2009, the CLEO Collaboration studied the semileptonic decays of $D^0\to K^-e^+\nu_e$ and $D^+\to \bar K^0e^+\nu_e$ decays by analyzing 818 pb$^{-1}$ data
collected at 3.773 GeV with the CLEO-c detector. Using double tag method, they measured the decay rates for semileptonic $D^0\to K^-e^+\nu_e$ and $D^+\to \bar K^0e^+\nu_e$ decays in nine $q^2$ bins~\cite{CLEOc}.
These measurements of decay rates are summarized in Tabs.~\ref{tab:Expt_D0} and \ref{tab:Expt_Dp}.

In 2006, the Belle Collaboration published the results on the $D^0\to K^-\ell^+\nu_\ell$ decays. They accumulated $56461\pm309\pm830$ inclusive $D^0$ mesons
and found $1318\pm37\pm7$ signal events for $D^0\to K^-e^+\nu_e$ decays and $1249\pm37\pm25$ signal events for $D^0\to K^-\mu^+\nu_\mu$ decays from a 282 fb$^{-1}$ data set collected around 10.58 GeV with the Belle detector~\cite{Belle}.
Using these selected events from semileptonic $D^0$ decays, they obtained the
form factors $f_+^K(q^2)$ in 27 $q^2$ bins with the bin size of 0.067 GeV$^2/c^4$.
To obtain the product $f_+^K(q^2_i)|V_{cs}|$ which will be used in our comprehensive analysis in Section~\ref{sec:ana},
we extrapolate these measurements of form factors at the Belle experiment to the product $f_+^K(q^2_i)|V_{cs}|$ using the PDG'2006 value of $|V_{cs}|=0.97296\pm0.00024$~\cite{pdg2006} which was originally used in the Belle's paper published.
Table~\ref{tab:ffVcs_Belle} lists the form factors $f_+^K(q^2_i)$ measured at the Belle experiment and our translated products $f_+^K(q^2_i)|V_{cs}|$.
These products will be used in our further analysis described in Section~\ref{sec:ana}.

\begin{table}[!hbp]
\caption{Measurements of form factors $f_+^K(q^2_i)$ at the Belle experiment and the products $f_+^K(q^2_i)|V_{cs}|$.
}
\label{tab:ffVcs_Belle}
\begin{center}
%\resizebox{0.48\textwidth}{!}{
\begin{tabular}{rccc}
\hline\hline
$i$ & $q^2_i$ (GeV$/c^2$) & $f_+^K(q^2_i)$ & $f_+^K(q^2_i)|V_{cs}|$ \\
\hline
 1 & 0.100 & $0.707\pm0.030$ & $0.688\pm0.029$ \\
 2 & 0.167 & $0.783\pm0.030$ & $0.762\pm0.029$ \\
 3 & 0.233 & $0.763\pm0.030$ & $0.743\pm0.029$ \\
 4 & 0.300 & $0.833\pm0.033$ & $0.811\pm0.032$ \\
 5 & 0.367 & $0.783\pm0.033$ & $0.762\pm0.032$ \\
 6 & 0.433 & $0.840\pm0.037$ & $0.817\pm0.036$ \\
 7 & 0.500 & $0.880\pm0.040$ & $0.856\pm0.039$ \\
 8 & 0.567 & $0.940\pm0.040$ & $0.915\pm0.039$ \\
 9 & 0.633 & $0.907\pm0.040$ & $0.882\pm0.039$ \\
10 & 0.700 & $0.820\pm0.040$ & $0.798\pm0.039$ \\
11 & 0.767 & $1.023\pm0.043$ & $0.996\pm0.042$ \\
12 & 0.833 & $0.997\pm0.047$ & $0.970\pm0.045$ \\
13 & 0.900 & $0.947\pm0.047$ & $0.921\pm0.045$ \\
14 & 0.967 & $1.043\pm0.053$ & $1.015\pm0.052$ \\
15 & 1.033 & $1.100\pm0.053$ & $1.070\pm0.052$ \\
16 & 1.100 & $0.937\pm0.057$ & $0.911\pm0.055$ \\
17 & 1.167 & $1.113\pm0.067$ & $1.083\pm0.065$ \\
18 & 1.233 & $1.097\pm0.070$ & $1.067\pm0.068$ \\
19 & 1.300 & $1.253\pm0.080$ & $1.219\pm0.078$ \\
20 & 1.367 & $1.380\pm0.087$ & $1.343\pm0.084$ \\
21 & 1.433 & $1.313\pm0.103$ & $1.278\pm0.101$ \\
22 & 1.500 & $1.190\pm0.110$ & $1.158\pm0.107$ \\
23 & 1.567 & $1.417\pm0.123$ & $1.378\pm0.120$ \\
24 & 1.633 & $1.473\pm0.173$ & $1.433\pm0.169$ \\
25 & 1.700 & $1.413\pm0.220$ & $1.375\pm0.214$ \\
26 & 1.767 & $1.147\pm0.340$ & $1.116\pm0.331$ \\
27 & 1.833 & $1.450\pm0.917$ & $1.411\pm0.892$ \\
\hline\hline
\end{tabular}%}
\end{center}
\end{table}
%%%%%%%%%%%%%%%%%%%%%%%%%%%%%%%%%%%%%%%%%%%%%%%%%%%%%%%%%%%%%%%%%%%%%%%%

\section{Analysis}
\label{sec:ana}

To obtain the product of the hadronic form factor at four momentum transfer $q=0$, $f^K_+(0)$, and the magnitude of CKM matrix element $|V_{cs}|$,
we perform a comprehensive $\chi^2$ fit to
these experimental measurements of the partial decay rates.
The object function to be minimized in the fit is defined as
\begin{equation}\label{eq:chi2}
    \chi^2 = \chi^2_{\rm R} + \chi^2_{\rm P} + \chi^2_{\rm F},
\end{equation}
where
$\chi^2_{\rm R}$ is for these measurements of decay branching fraction and/or partial decay rates in different $q^2$ ranges,
$\chi^2_{\rm P}$ corresponds to the products of $f_+^K(q^2_i)|V_{cs}|$ measured at the Belle experiment,
and
$\chi^2_{\rm F}$ is built for the measurements of $f_+^K(q^2_i)/f_+^K(0)$ measured at the FOCUS experiment.

Taking into account the correlations between these measurements, the quantity $\chi^2_{\rm R}$ is given by
\begin{equation}\label{eq:chi2_r}
    \chi^2_{\rm R} = \sum_{i=1}^{36}\sum_{j=1}^{36} (\Delta\Gamma^{\rm ex}_i - \Delta\Gamma^{\rm th}_i) (\mathcal C^{-1}_{\rm R})_{ij} (\Delta\Gamma^{\rm ex}_j - \Delta\Gamma^{\rm th}_j),
\end{equation}
where %$N=36$ is the total number of measured partial decay rates in $q^2$ bins at different experiments,
$\Delta\Gamma^{\rm ex}$ denotes the experimentally measured partial decay rate, $\Delta\Gamma^{\rm th}$ is the theoretical expectation of the decay rate, and
$\mathcal C^{-1}_{\rm R}$ is the inverse of the covariance matrix
$\mathcal C_{\rm R}$, which is a $36\times36$ matrix containing the correlations between the measured partial decay rates.
The construction of $C_{\rm R}$ is discussed in subsection~\ref{sec:cov}.
With the parametrization of the form factor, the theoretically predicted partial decay rate in a given $q^2$ bin is obtained by integrating Eq.~(\ref{eq:dG_dq2}) from the low boundary $q^2_{\rm low}$ to the up boundary $q^2_{\rm up}$ of the $q^2$ bin,
\begin{equation}\label{eq:DR_th}
    \Delta\Gamma^{\rm th} = \int_{q^2_{\rm low}}^{q^2_{\rm up}} \frac{G_F^2}{24\pi^3}|V_{cs}|^2 \boldsymbol{p}^3|f_+^K(q^2)|^2 dq^2.
\end{equation}
In this analysis, we used several forms of the form factor parameterizations which are discussed in subsection~\ref{sec:form_facor}.

Ignoring some possible correlations of the measurements of the product $f_+^K(q^2_i)|V_{cs}|$ measured at the Belle experiment,
the function $\chi^2_{\rm P}$ in Eq.~(\ref{eq:chi2}) is defined as
\begin{equation}\label{eq:chi2_P}
    \chi^2_{\rm P} = \sum_{i=1}^{27} \left( \frac{\tilde f_i^{\rm ex}-\tilde f_i^{\rm th}}{\sigma_i} \right)^2,
\end{equation}
where $\tilde f_i^{\rm ex}$ is the measured product $f_+^K(q^2)|V_{cs}|$ at the center of $i$th $q^2$ bin $q^2_i$ with the standard deviation $\sigma_i$, and $\tilde f_i^{\rm th}$ is the theoretical expectation of the product $f_+^K(q^2)|V_{cs}|$ at $q^2_i$.

Considering the correlations of the non-parametric form factors measured at the FOCUS experiment,
the $\chi^2_{\rm F}$ is constructed as
\begin{equation}\label{eq:chi2_F}
    \chi^2_{\rm F} = \sum_{i=1}^{9}\sum_{j=1}^{9} (F_i^{\rm ex}-F_i^{\rm th}) (\mathcal C^{-1}_{\rm F})_{ij} (F_j^{\rm ex}-F_j^{\rm th}),
\end{equation}
where $F_i^{\rm ex}$ is the measured relative form factor $f_+^K(q^2_i)/f_+^K(0)$
at $q^2_i$ from the FOCUS experiment, $F_i^{\rm th}$ is the theoretically expected value of $f_+^K(q^2_i)/f_+^K(0)$,
and $\mathcal C^{-1}_{\rm F}$ is the inverse of the covariance matrix
$\mathcal C_{\rm F}$. The construction of $\mathcal C_{\rm F}$ is described later in the subsection~\ref{sec:cov}.

\subsection{Form Factor Parameterizations}
\label{sec:form_facor}

In general, the single pole model is the simplest approach to describe the $q^2$ dependent behavior of form factor. The single pole model is expressed as
\begin{equation}\label{eq:ff_pole}
    f_+^K(q^2) = \frac{f_+^K(0)}{1-q^2/m_{\rm pole}^2},
\end{equation}
where $f_+^K(0)$ is the value of form factor at $q^2=0$, $m_{\rm pole}$ is the pole
mass which is predicted to be the mass of the $D^{*+}_s$ meson for semileptonic $D\to Ke^+\nu_e$ decays.

The so-called BK parameterization~\cite{BK} is also widely used in lattice QCD calculations and experimental studies of this decay. In the BK parameterization, the form factor
of the semileptonic $D\to Ke^+\nu_e$ decays is written as
\begin{equation}\label{eq:ff_BK}
    f_+^K(q^2) = \frac{f_+^K(0)}{(1-q^2/m_{D^{*+}_s}^2)(1-\alpha q^2/m_{D^{*+}_s}^2)},
\end{equation}
where $m_{D^{*+}_s}$ is the mass of the $D^{*+}_s$ meson,
and $\alpha$ is a free parameter to be fitted.
The value of $\alpha$ is assumed to be around $1.75$
for $D\to K\ell^+\nu_\ell$ in the BK parameterization.

The ISGW2 model~\cite{ISGW2} assumes
\begin{equation}\label{eq:ff_ISGW2}
    f_+^K(q^2) = f_+^K(q^2_{\rm max}) \left( 1+\frac{r^2}{12}(q^2_{\rm max} - q^2) \right)^{-2},
\end{equation}
where $q^2_{\rm max}$ is the kinematical limit of $q^2$,
and $r$ is the conventional radius of the meson.
In this model, the predictions of $f_+^K(q^2_{\rm max})$ and $r$ for $D\to K\ell^+\nu_\ell$ decays are $1.23$ and $1.12$ GeV$^{-1}c$, respectively.

The most general parameterization of the form factor is the series expansion~\cite{ff_zexpansion}, which is based on analyticity and unitarity.
In this parametrization, the variable $q^2$ is mapped to a new variable $z$ through
\begin{equation}
   z(q^2,t_0) = \frac{\sqrt{t_+-q^2}-\sqrt{t_+-t_0}}{\sqrt{t_+-q^2}+\sqrt{t_+-t_0}},
\end{equation}
with $t_{\pm}=(m_D\pm m_K)^2$ and $t_0 = t_+(1-\sqrt{1-t_-/t_+})$.
The form factor is then expressed in terms of the new variable $z$ as
\begin{equation}\label{eq:ff_series}
   f_+^K(q^2) = \frac{1}{P(q^2)\phi(q^2,t_0)} \sum_{k=0}^{\infty} a_k(t_0)[z(q^2,t_0)]^k,
\end{equation}
where $P(q^2)=z(q^2,m_{D^{*+}_s}^2)$ which accounts for the presence of the pole,
$\phi(q^2,t_0)$ is an arbitrary function,
and $a_k(t_0)$ are real coefficients.
In this analysis, the choice of $\phi(q^2,t_0)$ is taken to be
\begin{eqnarray}
  \phi(q^2,t_0) &=& \left( \frac{\pi m^2_c}{3} \right)^{\frac12} \left( \frac{z(q^2,0)}{-q^2} \right)^{\frac52} \left( \frac{z(q^2,t_0)}{t_0-q^2} \right)^{-\frac12}
  \nonumber
  \\
  &\times& \left( \frac{z(q^2,t_-)}{t_--q^2} \right)^{-\frac34} \frac{(t_+-q^2)}{(t_+-t_0)^{\frac14}},
\end{eqnarray}
where $m_c$ is the mass of charm quark, which is taken to be $1.2$ GeV$/c^2$.
In practical use, one usually make a truncation on the above series.
Actually, it is found that the current experimental data can be adequately described by only the first three
terms in Eq.~(\ref{eq:ff_series}).

In this analysis we will fit the measured decay rates to the three-parameter series expansion.
After optimizing the form factor parameters, we obtain the form for the three-parameter series expansion:
\begin{equation}\label{eq:ff_3series}
    f_+^K(q^2) =  \frac{f_+^K(0)P(0)\phi(0,t_0) (1+\sum_{k=1}^{2}r_k [z(q^2,t_0)]^k)}{P(q^2)\phi(q^2,t_0) (1+\sum_{k=1}^{2}r_k [z(0,t_0)]^k)},
\end{equation}
where $r_k\equiv a_k(t_0)/a_0(t_0)$ ($k=1,2$).

\subsection{Covariance Matrix}
\label{sec:cov}

It's a little complicated to compute the covariances of these 36 partial decay rates measured in different $q^2$ ranges and at different experiments. To be clear, we separate the correlations among these $\Delta\Gamma$ measurements into two case: the one associated with the experimental status of each independent experiment, and the one related to the external inputs of parameters such as the lifetime of the $D$ meson.

The statistical uncertainties in the $\Delta\Gamma$ measurements from the same experiment are correlated to some extent, while these are independent for the measurements from different experiments.
The systematic uncertainties from tracking, particle identification, etc. are usually independent between different experiments.
In this analysis, we treat the systematic uncertainties except the ones from $D$ lifetimes and branching fractions as fully uncorrelated between the measurements performed at different experiments.
We consider these below:

\begin{itemize}
\item
The covariances of the $\Delta\Gamma$ measured at the same experiment are computed using the statistical errors,  the systematic errors, and the correlation coefficients, which are presented in their original papers published.

\item
For the measurements of $D^0\to K^-e^+\nu_e$ decay, the lifetime of $D^0$ meson is used to obtain the partial decay rates in particular $q^2$ ranges. The systematic uncertainties due to imperfect knowledge of $D^0$ lifetime are fully correlated among all these measurements of the partial rates of $D^0\to K^-e^+\nu_e$ decay.
Similarly, the systematic uncertainties related to $D^+$ lifetime are fully correlated among all of the $\Delta\Gamma$ measurements for $D^+\to\bar K^0e^+\nu_e$ decay.

\item
An additional systematic uncertainty from $B(D^0\to K^-\pi^+)$ is fully correlated between these relative measurements of $D^0\to K^-e^+\nu_e$ decay at the E691, CLEO, CLEO-II and BaBar experiments.
Since we only use one relative measurement of $D^+\to \bar K^0e^+\nu_e$ decay which is from the CLEO-II experiment,
there are no correlations due to the normalization branching fraction $B(D^+\to\bar K^0\pi^+)$ between this measurement and other measurements.
\end{itemize}

With these considerations mentioned above, we then construct a $36\times36$ covariance matrix $\mathcal C_{\rm R}$ which is necessary in  the form factor fit.

The entry in $i$-th row and $j$-th column of the covariance matrix $\mathcal C_{\rm R}$ in Eq.~(\ref{eq:chi2_F}) is given by $(\mathcal C_{\rm R})_{ij}=\sigma_i\sigma_j\rho_{ij}$,
where $\sigma_i$ and $\sigma_j$ are the errors of the  $f_+^K(q^2)/f_+^K(0)$ at $q^2_i$ and $q^2_j$ measured at the FOCUS experiment, respectively, and $\rho_{ij}$ is the correlation coefficient of these two measurements of $f_+^K(q^2)/f_+^K(0)$.
The values of the errors and correlation coefficients are directly quoted from Ref.~\cite{FOCUS}.

\subsection{Fits to Experimental Data}

Four fits are applied to the experimental data with the form factor hypothesis of single pole model, modified pole model, ISGW2 model and series expansion.
The fit to experimental data returns the normalization $f_+^K(0)|V_{cs}|$ and the shape parameters of the form factor which govern the behavior of form factor in high $q^2$ range.

\begin{figure*}
  \includegraphics[width=\textwidth]{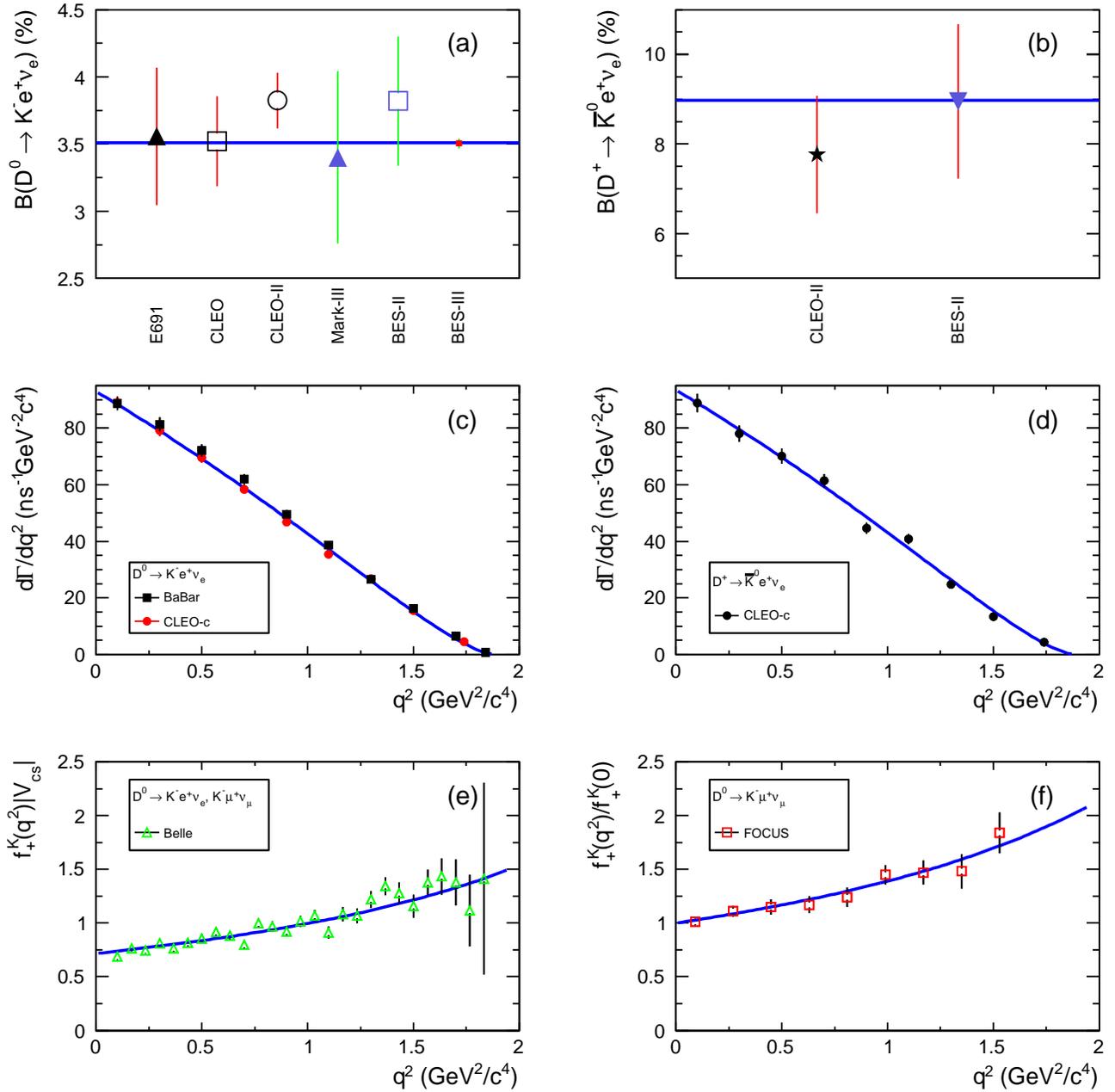}
  \caption{
  Comparisons of branching fraction measurements for (a) $D^0\to K^-e^+\nu_e$, (b) $D^+\to \bar K^0e^+\nu_e$,
  (c) measurements of differential decay rates for $D^0\to K^-e^+\nu_e$ measured at the BaBar and CLEO-c experiments, (d) differential decay rates for $D^+\to \bar K^0e^+\nu_e$ measured at the CLEO-c experiment, (e) the product of form factor and $|V_{cs}|$ measured at the Belle experiment, and (f) the normalized form factor measured at the FOCUS experiment. The blue lines show the fit to these measurements using the series expansion for the form factor.
  }
  \label{fig:fit}
\end{figure*}

The numerical results of the fit corresponding to each form of the form factor parameterization are summarized in Tab.~\ref{tab:results}.
As an example, figure~\ref{fig:fit}
presents the result of the fit in the case of using the form factor parameterization of series expansion.
In Fig.~\ref{fig:fit} (a) and (b), we compared the measured branching fractions of $D^0\to K^-e^+\nu_e$ and $D^+\to \bar K^0e^+\nu_e$ decays from different experiments.
Figure~\ref{fig:fit} (c) and (d) show the measured differential decay rates for $D^0\to K^-e^+\nu_e$ and $D^+\to \bar K^0e^+\nu_e$, respectively.
Figure~\ref{fig:fit} (e) depicts the measurements of $f_+^K(q^2)|V_{cs}|$ at different $q^2$ from the Belle experiments.
The FOCUS measurements of the normalized form factor $f_+^K(q^2)/f_+^K(0)$ are illustrated in Fig.~\ref{fig:fit} (f).
In these figures, the lines show the best fit to these measurements.

%%%%%%%%%%%%%%%%%%%%%%%%%%%%%%%%%%%%%%%%%%%%%%%%%%

\begin{table*}
  \centering
  \caption{Fitted parameters corresponding to different form factor parameterizations and $\chi^2/{\rm d.o.f.}$ of the fit.}
  \label{tab:results}
  \begin{tabular}{lccc}
  \hline\hline
  Parameterization & $f_+^K(0)|V_{cs}|$ & Shape parameters &  $\chi^2/{\rm d.o.f.}$ \\
  \hline
  Single pole & $0.720 \pm 0.003$ & $M_{\rm pole}=(1.909 \pm 0.011)$ GeV$/c^2$ & $106.0/70$ \\
  BK & $0.716 \pm 0.003$ & $\alpha = 0.327 \pm 0.021$ &  $101.0/70$ \\
  ISGW2 & $0.714 \pm 0.003$ & $r=(1.610 \pm 0.015)$ GeV$^{-1}c^2$ & $101.9/70$ \\
  Series expansion & $0.717 \pm 0.004$ & $r_1=-2.34 \pm 0.17$  & $101.1/69$ \\
  & & $r_2=\phantom{-}0.43 \pm 3.82$ &  \\
  \hline\hline
  \end{tabular}
\end{table*}
%%%%%%%%%%%%%%%%%%%%%%%%%%%%%%%%%%%%%%%%%%%%%%%%%%

\begin{figure*}
  \includegraphics[width=\textwidth]{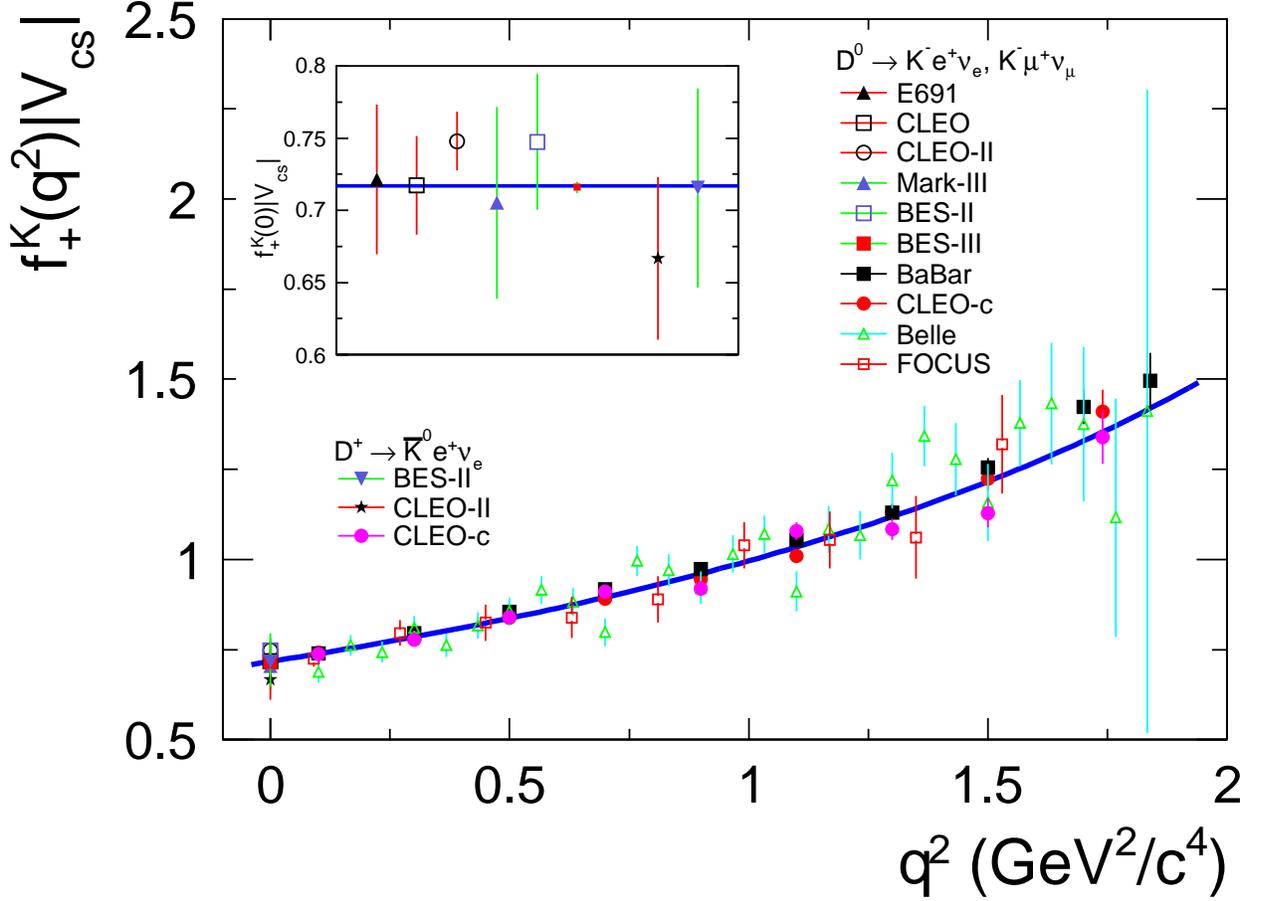}
  \caption{The product $f_+^K(q^2)|V_{cs}|$ measured at different experiments as a function of $q^2$. The blue curve represents the series expansion fit to these $f_+^K(q^2)|V_{cs}|$. The insert plot shows the comparison of the products $f_+^K(0)|V_{cs}|$ which are obtained using the branching fractions measured at different experiments. }
  \label{fig:ff_x_Vcs}
\end{figure*}
%%%%%%%%%%%%%%%%%%%%%%%%%%%%%%%%%%%%%%%%%%%%%%%%%%

To check the fit quality and also the isospin invariance, the experimentally  measured decay branching fractions and/or partial rates are mapped into the product $f_+^K(q^2_i)|V_{cs}|$ via
\begin{equation}\label{eq:f0_B}
    f_+^K(0)|V_{cs}| = \sqrt{\frac{B}{\tau_D}\frac{1}{N}}
\end{equation}
and
\begin{equation}\label{eq:f0_DR}
    f_+^K(q^2_i)|V_{cs}| = \sqrt{\left(\frac{d\Gamma}{dq^2}\right)_i\frac{24\pi^3}{G_F^2\boldsymbol p_i^3}},
\end{equation}
where
$B$ denotes the measured branching fraction, the differential decay rate $(d\Gamma/dq^2)_i$ is obtained by dividing measured decay rate in $q^2$ bin $i$ by the corresponding bin size.
The normalization $N$ is given by
\begin{equation}\label{eq:norm}
    N = \frac{G_F^2}{24\pi^3|f_+^K(0)|^2}\int_{0}^{q^2_{\rm max}} {\boldsymbol p}^3|f_+^K(q^2)|^2dq^2.
\end{equation}
The effective $\boldsymbol p_i^3$ in $q^2$ bin $i$ is given by
\begin{equation}\label{eq:p3}
    \boldsymbol p_i^3 = \frac{\int_{q^2_{\rm low}}^{q^2_{\rm up}} {\boldsymbol p}^3|f_+^K(q^2)|^2dq^2}{|f_+^K(q^2_i)|^2(q^2_{\rm up}-q^2_{\rm low})}.
\end{equation}
To calculate the integral in Eqs.~(\ref{eq:norm}) and (\ref{eq:p3}), we use the shape parameters of form factor, which is obtained from the series expansion fit to the data.

Figure~\ref{fig:ff_x_Vcs} shows the product $f_+^K(q^2)|V_{cs}|$ as a function of $q^2$,
where
the blue curve corresponds to the best series expansion fit to the experimental data.
In this fit, eight measurements of $f_+^K(0)|V_{cs}|$ locate at $q^2=0$, which overlap each other.
To be clear, these $f_+^K(0)|V_{cs}|$ translated from the decay branching fractions measured at different experiments are also displayed in the insert plot in Fig.~\ref{fig:ff_x_Vcs}.

\section{Results}
\label{sec:rslt}

In this analysis, we choose the result from the fit using series expansion as our primary results and use this to extract the form factor $f_+^K(0)$ and the magnitude of the CKM matrix element $V_{cs}$.

\subsection{Form Factor $f_+^K(0)$}

\begin{figure}[h]
  \includegraphics[width=0.5\textwidth]{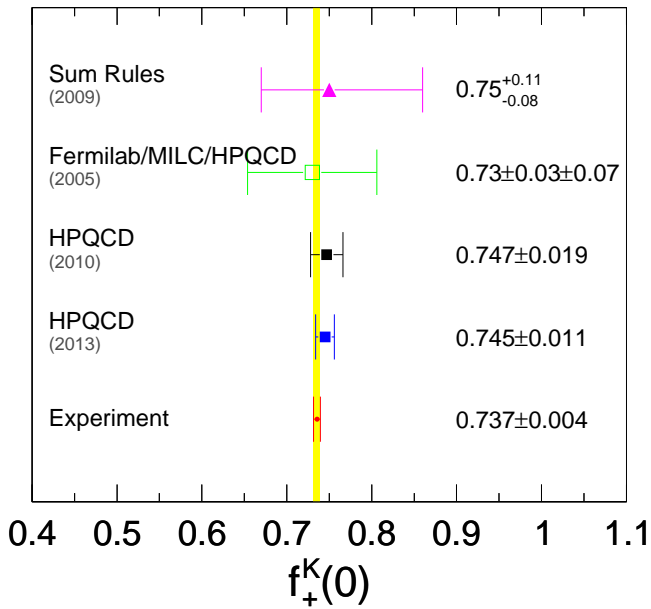}
  \caption{Comparison of our determined form factor from experimental measurements with the theoretical calculations of the form factor. }
  \label{fig:Cmp_f0}
\end{figure}

Dividing the value of $f_+^K(0)|V_{cs}|=0.717\pm0.004$ shown in Tab.~\ref{tab:results} from the series expansion fit
by the
$|V_{cs}|=0.97343\pm0.00015$ obtained using unitarity constraints~\cite{pdg}
yields the form factor
\begin{equation}\label{eq:f0}
   f_+^K(0) = 0.737\pm0.004\pm0.000,
\end{equation}
where the first uncertainty is from the combined statistical and systematic uncertainties in the partial decay rate measurements,
and the second is due to the uncertainty in the $|V_{cs}|$.
The result for the form factor determined in this analysis is compared with the theoretical calculations of the form factor from the lattice QCD ~\cite{HPQCD, HPQCD2010, Fermilab2005}
and from QCD light-cone sum rules~\cite{SR} in Fig.~\ref{fig:Cmp_f0}.
Our result of the form factor extracted by analyzing all existing experimental measurements is consistent with these values predicted by theory, but is with higher precision than the most accurate one from LQCD calculation by a factor of 2.8.

\subsection{Parameters of Form Factor}

\begin{figure}
  \includegraphics[width=0.5\textwidth]{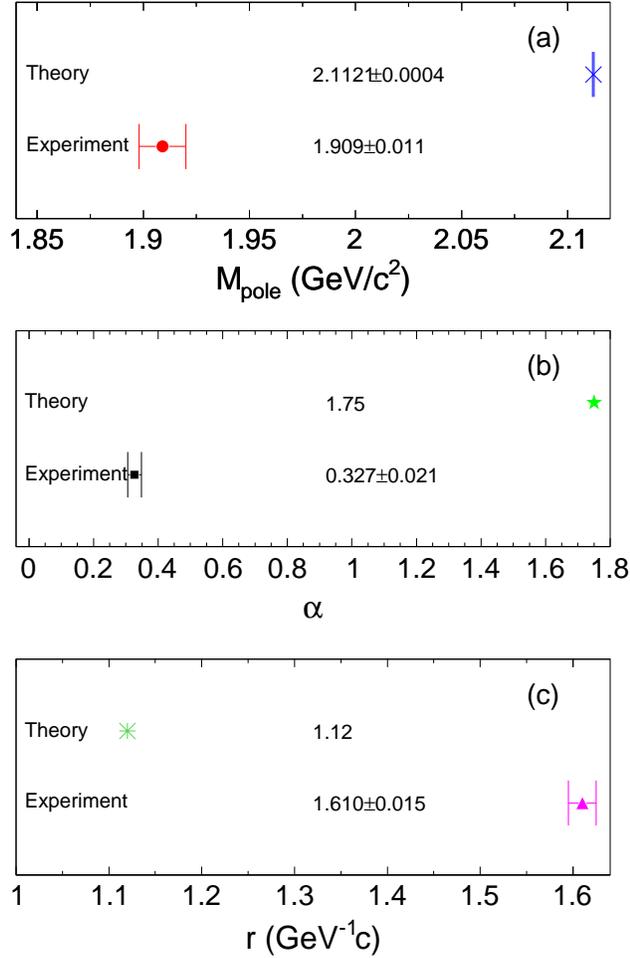}
  \caption{Comparisons of the form factor parameters determined from experimental measurements and the theoretical expectations: (a) the pole mass $M_{\rm ploe}$ in single pole model, (b) $\alpha$ in the BK model, and (c) $r$ in the ISGW2 model. }
  \label{fig:Cmp_par}
\end{figure}

When these shape parameters of the form factor parameterization are left free in the fit, the form factor parametrizations of the single pole model, BK model, the ISGW2 model, and the series expansion model are all capable of describing the experimental data with almost identical $\chi^2$ probability.
However, for the physical interpretation of the shape parameters in the single pole model, BK model, the ISGW2 model, the values of the parameters obtained from the fits
are largely deviated from those expected values by these models.
This indicates that the experimental data do not support the physical interpretation of the shape parameters in these parametriziations.
Figure~\ref{fig:Cmp_par} (a), (b) and (c) show the comparisons between the measured values and the theoretically expected values for the pole mass $M_{\rm pole}$ in single pole model, $\alpha$ in BK model, and $r$ in ISGW2 model.
These measured parameters do not agree with the values predicted by these form factor models.

\subsection{CKM Matrix Element $|V_{cs}|$}

\begin{figure}[h]
  \includegraphics[width=0.5\textwidth]{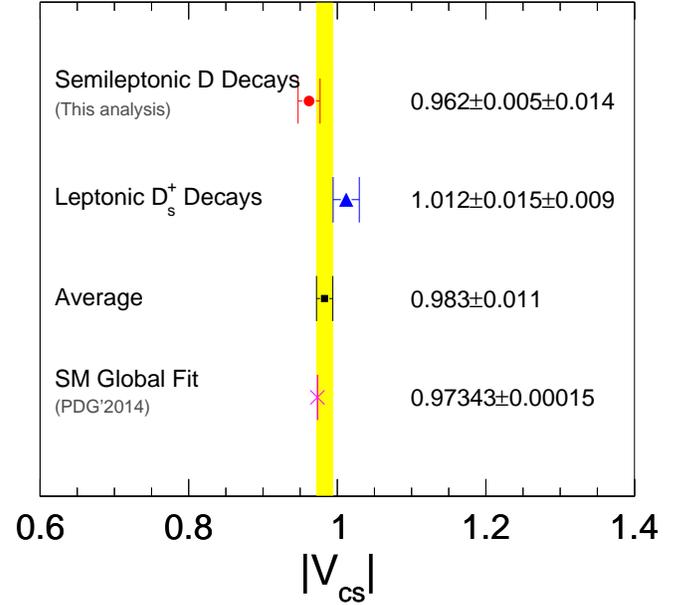}
  \caption{Comparison of $|V_{cs}|$ extracted from semileptonic $D$ decays in this analysis with the one extracted from leptonic $D_s^+$ decays and the one from SM global fit.}
  \label{fig:Cmp_Vcs}
\end{figure}

\begin{figure}[h]
  \includegraphics[width=0.5\textwidth]{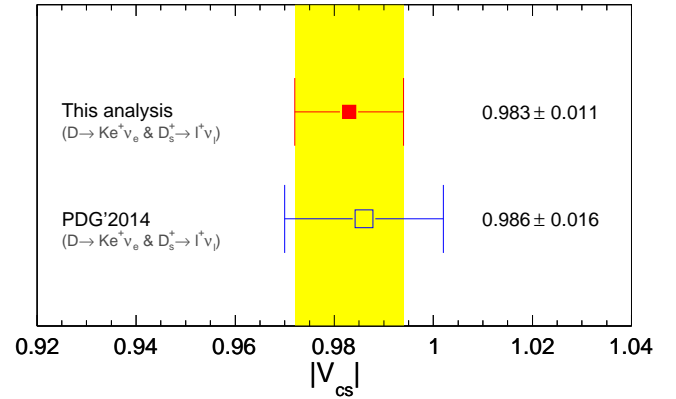}
  \caption{Comparison of $|V_{cs}|$ extracted from semileptonic $D$ decays and leptonic $D^+_s$ decays in this analysis with the PDG value.}
  \label{fig:Cmp_Vcs_PDG}
\end{figure}

Using the product $f_+^K(0)|V_{cs}|=0.717\pm0.004$ obtained from the comprehensive series expansion fit in conjunction with the form factor $f_+^K(0)=0.745\pm0.011$~\cite{HPQCD} calculated in LQCD for the $D\to K$ transition,
we determine the magnitude of the CKM matrix element $V_{cs}$ to be
\begin{equation}
|V_{cs}|^{D\to Ke^+\nu_e} = 0.962\pm0.005\pm0.014,
\end{equation}
where the last uncertainty corresponds to the accuracy of the form factor $f^K_+(0)$
calculated in LQCD.
Combining with the value $|V_{cs}|^{D^+_s\to\ell^+\nu_\ell}=1.012\pm0.015\pm0.009$, which is extracted from the measurements of leptonic $D_s^+$ decays (see Appendix~\ref{sec:ap}),
we obtain the magnitude of the CKM matrix element $V_{cs}$ to be
\begin{equation}
|V_{cs}| = 0.983\pm0.011.
\end{equation}
Figure~\ref{fig:Cmp_Vcs} shows the comparisons of the value of $|V_{cs}|$ which is determined
with the $|V_{cs}|^{D\to Ke^+\nu_e}$ in this analysis together with the $|V_{cs}|^{D^+_s\to\ell^+\nu_\ell}$ determined from leptonic $D_s^+$ decays,
and the value from a SM global fit~\cite{pdg}.
Figure~\ref{fig:Cmp_Vcs_PDG} shows a comparison of our extracted $|V_{cs}|$
from all existing measurements of $D\to Ke^+\nu_e$ and $D^+_s\to\ell^+\nu_\ell$ decays
along with the PDG'2014 value of the $|V_{cs}|$ determined with CLEO-c, BaBar and Belle's measurements
of $D\to Ke^+\nu_e$ and $D^+_s\to\ell^+\nu_\ell$ decays~\cite{pdg}.

The $|V_{cs}|^{D\to Ke^+\nu_e}$ extracted from semileptonic $D$ decays deviates from the
$|V_{cs}|^{D^+_s\to\ell^+\nu_\ell}$ extracted from leptonic $D_s^+$ decays by $2.2\sigma$.
This discrepancy may arise from
three sources: 1) some new physic effects involved in leptonic $D_s^+$ decays,
which modify the decay rate; 2) underestimated decay constant $f_{D^+_s}$ in LQCD;
3) overestimated form factor $f_+^K(0)$ in LQCD.
Any of these would modify these decay rates resulting in shifts of the $|V_{cs}|^{D\to Ke^+\nu_e}$
and $|V_{cs}|^{D^+_s\to\ell^+\nu_\ell}$,
which are extracted from semileptonic $D$ and leptonic $D^+_s$ decays, respectively.

\subsection{Effects of Radiative Correction on Decay Rate}

From experimental aspect, it's difficult to exclude the soft photon emission in the final state in the
procedure of event selection.
As a consequence,
the experimentally measured branching fractions or partial decay rates usually include the contribution
from the soft photon emission in the final states more or less.
The theoretical prediction for the decay rate,
Eq. (\ref{eq:dG_dq2}), used in the extractions of form factor $f_+^K(0)$ and
CKM matrix element $|V_{cs}|$
is for the tree-level semileptonic $D$ decay process.
As the experimental precision of the decay rate measurement has already achieved
an accuracy level of $0.5\%$,
in addition to improve the precision of the LQCD calculation for $f_+^K(0)$,
it should be also necessary to take into account the radiative correction
in further precise determination of $|V_{cs}|$.
However, unlike the situation in $K_{\ell3}$ decays, due to the lack of an universally valid effective theory,
the theoretical or phenomenological estimation of
radiative corrections in the semileptonic $D$ decays is absent at present stage.
So we ignore this radiative correction in the analysis at present.

\section{Summary}
\label{sec:sum}

By globally analyzing all existing branching fractions of the $D\to Ke^+\nu_e$ decays measured at earlier experiments and recent BESIII experiment as well as the precise measurements of partial decay rates in $q^2$ bins performed at the BaBar and CLEO-c experiments together, we  obtain the most precise product of form factor and the magnitude of CKM matrix element $V_{cs}$ from a comprehensive $\chi^2$ fit.
This obtained product reflects all of measurements for $D\to Ke^+\nu_e$ decays in the world in the last 25 years.
With the obtained $f_+^K(0)|V_{cs}|$ in conjunction with $|V_{cs}|$ from SM global fit, we determined the form factor
\begin{equation*}
    f_+^K(0) = 0.737\pm0.004\pm0.000,
\end{equation*}
which is in good agreement within error with LQCD calculations, but more precise than the most accurate LQCD calculation of the form factor by 2.8 factors.
Alternately, with the recent most precise semileptonic $D\to K e^+\nu_e$ decay form factor calculated in LQCD,
we obtain the $|V_{cs}|^{D\to Ke^+\nu_e}=0.962\pm0.005\pm0.014$, where the error is dominated by the uncertainties in LQCD calculation of the hadronic form factor.
This determined $|V_{cs}|$ is in good agreement within error with the one from SM global fit, which indicates that no evidence of new physic effects involved in the semileptonic $D\to Ke^+\nu_e$ decays is observed at present experimental accuracy level.

If combining the $|V_{cs}|^{D\to Ke^+\nu_e} = 0.962\pm0.005\pm0.014$ determined from semileptonic $D$ decays and $|V_{cs}|^{D^+_s\to\ell^+\nu_\ell} = 1.012\pm0.015\pm0.009$ determined from leptonic $D_s^+$ decays together, we find
\begin{equation*}
    |V_{cs}| = 0.983\pm0.011,
\end{equation*}
which improves the accuracy of the PDG'2014 value $|V_{cs}|^{\rm PDG'2014}=0.986\pm0.016$ by $45\%$,
and is the most precisely extracted $|V_{cs}|$ from all existing measurements of semileptonic $D$ decays
and leptonic $D_s^+$ decays up to date.

\section*{Acknowledgements}
This work is supported in part by the Ministry of Science of Technology of China under Contracts No. 2009CB825204; National Natural Science Foundation of China (NSFC) under Contacts No. 10935007 and No. 11305180.

\appendix
\section{Extraction of $|V_{cs}|$ from Leptonic $D_s^+$ Decays}
\label{sec:ap}

In this appendix, we present the determination of $|V_{cs}|$ by analyzing the existing measurements of leptonic $D_s^+\to\ell^+\nu_\ell$ ($\ell=\mu,\tau$) decays.

In SM of particle physics, the decay width of $D_s^+\to\ell^+\nu_\ell$ is given by
\begin{equation}\label{eq:DR_lep}
    \Gamma(D_s^+\to\ell^+\nu_\ell) = \frac{G_F^2}{8\pi} m_\ell^2 m_{D_s^+} \left(1 - \frac{m_\ell^2}{m_{D_s^+}^2} \right)^2 f_{D_s^+}^2 |V_{cs}|^2,
\end{equation}
where $m_\ell$ is the mass of lepton and $m_{D_s^+}$ is the mass of $D_s^+$ meson.
The parameter $f_{D_s^+}$ is the decay constant, which is associated with the strong interaction effects between the two initial-state quarks.

In the past two decades, many measurements of leptonic $D_s^+\to\ell^+\nu_\ell$ decays were performed at $e^+e^-$ experiments and fixed-target experiments.  These measured branching fractions are summarized in Tab.~\ref{tab:BF_Dstoln}.

\begin{table}[h]
  \centering
  \caption{Measurements of $B(D_s^+\to\mu^+\nu_\mu)$ and $B(D_s^+\to\tau^+\nu_\tau)$.}
  \label{tab:BF_Dstoln}
  %\resizebox{0.48\textwidth}{!}{
  \begin{tabular}{lc}
    \hline
    \hline
    Experiment & $B(D_s^+\to\mu^+\nu_\mu)$ (\%) \\
    %\hline
    BES-I~\cite{BESI_leptonic}  & $1.5^{+1.3+0.3}_{-0.6-0.2}$ \\
    ALEPH~\cite{aleph} & $0.68\pm0.11\pm0.18$ \\
    CLEO-c~\cite{CLEOc_munu} & $0.565\pm0.045\pm0.017$ \\
    BaBar~\cite{BaBar_Dstolnu} & $0.602\pm0.038\pm0.034$ \\
    Belle~\cite{Belle_Dstolnu} & $0.531\pm0.028\pm0.020$ \\
    \hline
    %Experiment & $\frac{\Gamma(D_s^+\to\mu^+\nu_\mu)}{\Gamma(D_s^+\to\phi\pi^+)}$ \\
    Experiment & $\Gamma(D_s^+\to\mu^+\nu_\mu) / \Gamma(D_s^+\to\phi\pi^+)$ \\
    BEATRICE~\cite{BEATRICE} & $0.23\pm0.06\pm0.04$ \\
    CLEO-II~\cite{CLEOII_munu} & $0.173\pm0.023\pm0.035$ \\
    BaBar~\cite{BaBar_munu} & $0.143\pm0.018\pm0.006$ \\
    \hline
    Experiment & $B(D_s^+\to\tau^+\nu_\tau)$ (\%) \\
    L3~\cite{L3} & $7.4\pm2.8\pm2.4$ \\
    OPAL~\cite{opal} & $7.0\pm2.1\pm2.0$ \\
    ALEPH~\cite{aleph} & $5.79\pm0.77\pm1.84$ \\
    CLEO-c~\cite{CLEOc_taunu} & $5.58\pm0.33\pm0.13$ \\
    BaBar~\cite{BaBar_Dstolnu} & $5.00\pm0.35\pm0.49$ \\
    Belle~\cite{Belle_Dstolnu} & $5.70\pm0.21^{+0.31}_{-0.30}$ \\
    \hline
    \hline
  \end{tabular}%}
\end{table}

To extract the magnitude of CKM matrix element $V_{cs}$,
we globally analyze all of these existing measurements of leptonic $D^+_s$ decay branching fractions shown in Tab.~\ref{tab:BF_Dstoln}.
Assuming lepton universality,
we construct a object function $\chi^2$:
\begin{eqnarray}\label{eq:chi2_Dstoln}
  \chi^2 &=& \sum_{i=1}^{5} \left( \frac{B_{\mu,i}^{\rm ex} - B_{\mu}^{\rm th}}{\sigma_{i}} \right)^2
 + \sum_{j=1}^{3} \left( \frac{R_{\mu,j}^{\rm ex} - R_{\mu}^{\rm th}}{\sigma_{j}} \right)^2
 \nonumber
 \\
   &+& \sum_{k=1}^{6} \left( \frac{B_{\tau,k}^{\rm ex} - B_{\tau}^{\rm th}}{\sigma_{k}} \right)^2,
\end{eqnarray}
where
$B_{\mu,i}^{\rm ex}$ is the $i$th experimentally measured branching fraction of $D_s^+\to\mu^+\nu_\mu$ decay,
$R_{\mu,j}^{\rm ex}$ is the $j$th experimentally measured partial width of $D_s^+\to\mu^+\nu_\mu$ decay relative to the partial width of $D_s^+\to\phi\pi^+$ decay,
$B_{\tau,k}^{\rm ex}$ is the $k$th experimentally measured branching fraction of $D_s^+\to\tau^+\nu_\tau$ decay,
and $\sigma$ denotes the combined statistical and systematic error of the measured (relative) branching fraction.
The theoretically predicted branching fraction of $D_s^+\to\mu^+\nu_\mu$ is given by
\begin{equation}\label{BF_Dstomunu}
   B_{\mu}^{\rm th} = \Gamma(D_s^+\to\mu^+\nu_\mu) \times \tau_{D^+_s},
\end{equation}
where
$\Gamma(D_s^+\to\mu^+\nu_\mu)$ is calculated with Eq.~(\ref{eq:DR_lep}),
and $\tau_{D^+_s}$ is the lifetime of $D_s^+$ meson.
The theoretically expected ratio of $\Gamma(D_s^+\to\mu^+\nu_\mu)$ over $\Gamma(D_s^+\to\phi\pi^+)$ is given by
\begin{equation}\label{R_Dstomunu}
   R_\mu^{\rm th} = B_{\mu}^{\rm th} / B(D_s^+\to\phi\pi^+),
\end{equation}
where $B(D_s^+\to\phi\pi^+)$ is the branching fraction of $D_s^+\to\phi\pi^+$ decay.
The theoretically predicted branching fraction of $D_s^+\to\tau^+\nu_\tau$ is given by
\begin{equation}\label{BF_Dstomunu}
   B_{\tau}^{\rm th} = \Gamma(D_s^+\to\tau^+\nu_\tau) \times \tau_{D^+_s},
\end{equation}
where
$\Gamma(D_s^+\to\tau^+\nu_\tau)$ is calculated with Eq.~(\ref{eq:DR_lep}).

To obtain the experimentally measured product $f_{D_s^+}|V_{cs}|$, we perform a $\chi^2$ fit to these measured branching fractions for $D^+_s\to\ell^+\nu_\ell$ decays shown in Tab.~\ref{tab:BF_Dstoln}.
In the fit, we use
$m_\mu=(105.6583715\pm0.0000035)$ MeV,
$m_\tau=(1776.82\pm0.16)$ MeV,
$m_{D_s^+}=(1968.30\pm0.11)$ MeV,
$\tau_{D_s^+}=(500\pm7)\times10^{-15}$ s,
and
$B(D_s^+\to\phi\pi^+)=(4.5\pm0.4)\%$
which are all quoted from PDG~\cite{pdg}.
The product of the decay constant and the magnitude of CKM matrix element $V_{cs}$ is the only free parameter in the fit.
The fit returns
\begin{equation}\label{eq:fDsVcs}
    f_{D_s^+}|V_{cs}|=(252.0 \pm 3.7 \pm 1.8)~\rm MeV,
\end{equation}
where the first error is from the statistical and systematic uncertainties in the measured (relative) branching fractions,
and the second error is due to the uncertainties in the masses of lepton and $D_s^+$ meson, the lifetime of $D^+_s$ meson, and the branching fraction of $D^+_s\to\phi\pi^+$ decay.

Dividing the product $f_{D_s^+}|V_{cs}|$ by the value $f_{D_s^+}=(249.0\pm0.3^{+1.1}_{-1.5})$ MeV
which is the newest and most precise value of decay constant calculated in LQCD with $N_f=2+1+1$ quark flavors~\cite{LQCD_fDs},
we obtain
\begin{equation}
    |V_{cs}|^{D^+_s\to\ell^+\nu_\ell}=1.012\pm0.015\pm0.009,
\end{equation}
where the first error is from the statistical and systematic uncertainties in the measured (relative) branching fractions,
and the second error is mainly due to the uncertainties in the lifetime of $D^+_s$ meson, and the $f_{D_s^+}$ calculated in lattice QCD.

Alternatively, by inserting $|V_{cs}|=0.97343\pm0.00015$ from the SM global fit~\cite{pdg} into Eq.~(\ref{eq:fDsVcs}), we determine
\begin{equation}
    f_{D^+_s}=(258.9\pm3.8\pm1.8)~\rm MeV,
\end{equation}
which is the most precisely determined $D^+_s$ leptonic decay constant.


\begin{thebibliography}{99}

\bibitem{RongG_program_ccbar}
 G. Rong, Chin. Phys. C {\bf 34}, 788 (2010).

\bibitem{E691}
 J. C. Anjos {\it et al.} (The Tagged Photon Spectrometer Collaboration), Phys. Rev. Lett. {\bf 62}, 1587 (1989).

\bibitem{CLEO}
 G. Crawford {\it et al.} (CLEO Collaboration), Phys. Rev. D {\bf 44}, 3394 (1991).

\bibitem{CLEOII}
 A. Bean {\it et al.} (CLEO Collaboration), Phys. Lett. B {\bf 317}, 647 (1993).

\bibitem{BaBar}
 B. Aubert {\it et al.} (BaBar Collaboration), Phys. Rev. D {\bf 76}, 052005 (2007).

\bibitem{B_D0toKpi_BaBar}
 B. Aubert {\it et al.} (BaBar Collaboration), Phys. Rev. Lett. {\bf 100}, 051802 (2008).

\bibitem{B_D0toKpi_CLEOc}
 S. Dobbs {\it et al.} (CLEO Collaboration), Phys. Rev. D {\bf 76}, 112001 (2007).

\bibitem{B_D0toKpi_CLEOII}
 M. Artuso {\it et al.} (CLEO Collaboration), Phys. Rev. Lett. {\bf 80}, 3193 (1998).

\bibitem{B_D0toKpi_ALEP}
 R. Barate {\it et al.} (ALEPH Collaboration), Phys. Lett. B {\bf 403}, 367 (1997).

\bibitem{B_D0toKpi_ALEP2}
 D. Decamp {\it et al.} (ALEPH Collaboration), Phys. Lett. B {\bf 266}, 218 (1991).

\bibitem{B_D0toKpi_ARG}
 H. Albrecht {\it et al.} (ARGUS Collaboration), Phys. Lett. B {\bf 340}, 125 (1994).

\bibitem{B_DptoKLpi_CLEOc}
 Q. He {\it et al.} (CLEO Collaboration), Phys. Rev. Lett. {\bf 100}, 091801 (2008).

\bibitem{pdg}
  K. A. Olive {\it et al.} (Particle Data Group), Chin. Phys. C {\bf 38}, 090001 (2014).

\bibitem{FOCUS}
 J. M. Link {\it et al.} (FOCUS Collaboration), Phys. Lett. B {\bf 607}, 233 (2005).

\bibitem{MarkIII}
  J. Adler {\it et al.} (Mark III Collaboration), Phys. Rev. Lett. {\bf 62}, 1821 (1989).

\bibitem{BESII_D0}
  M. Ablikim {\it et al.} (BES Collaboration), Phys. Lett. B {\bf 597}, 39 (2004).

\bibitem{BESII_Dp}
  M. Ablikim {\it et al.} (BES Collaboration), Phys. Lett. B {\bf 608}, 24 (2005).

\bibitem{BESIII_D0Kenu}
  Y. H. Zheng (For BESIII Collaboration), ICHEP2014, 2-7 July 2014, Valencia (Spain). \\
  H. L. Ma (For BESIII Collaboration), Beauty2014, 14-18 July 2014, Edinburgh (UK).
  \\
  G. Rong (For BESIII Collaboration), CKM2014, 8-12 September 2014, Vienna (Austria).

\bibitem{CLEOc}
  D. Besson {\it et al.} (CLEO Collaboration), Phys. Rev. D {\bf 80}, 032005 (2009).

\bibitem{Belle}
  L. Widhalm {\it et al.} (Belle Collaboration), Phys. Rev. Lett. {\bf 97}, 061804 (2006), arXiv:hep-ex/0604049.

\bibitem{pdg2006}
  W.-M. Yao {\it et al.}, Journal of Physics G {\bf 33}, 1 (2006).

\bibitem{BK}
  D. Becirevcic and A. B. Kaidalov, Phys. Lett. B {\bf 478}, 417 (2000).

\bibitem{ISGW2}
  D. Scora and N. Isgur, Phys. Rev. D {\bf 52}, 2783 (1995).

\bibitem{ff_zexpansion}
  T. Becher and R. J. Hill, Phys. Lett. B {\bf 633}, 61 (2006).

\bibitem{HPQCD}
   J. Koponen {\it et al.} (HPQCD Collaboration), arXiv: 1305.1462 [hep-lat].

\bibitem{HPQCD2010}
  H. Na {\it et al.} (HPQCD Collaboration), Phys. Rev. D {\bf 82}, 114506 (2010).

\bibitem{Fermilab2005}
  C. Aubin {\it et al.} (Fermilab Lattice Collaboration, MILC Collaboration, and HPQCD Collaboration), Phys. Rev. Lett. {\bf 94}, 011601 (2005).

\bibitem{SR}
  A. Khodjamirian {\it et al.}, Phys. Rev. D {\bf 80}, 114005 (2009).

\bibitem{BESI_leptonic}
  J. Z. Bai {\it et al.} (BES Collaboration), Phys. Rev. Lett. {\bf 74}, 4599 (1995).

\bibitem{aleph}
  A. Heister {\it et al.} (ALEPH Collaboration), Phys. Lett. B {\bf 528}, 1 (2002).

\bibitem{CLEOc_munu}
  J. P. Alexander {\it et al.} (CLEO Collaboration), Phys. Rev. D {\bf 79}, 052001 (2009).

\bibitem{BaBar_Dstolnu}
  P. del Amo Sanchez {\it et al.} (BaBar Collaboration), Phys. Rev. D {\bf 82}, 091103(R) (2010).

\bibitem{Belle_Dstolnu}
  A. Zupanc {\it et al.} (Belle Collaboration), JHEP {\bf 09}, 139 (2013).

\bibitem{BEATRICE}
  Y. Alexandrov {\it et al.} (BEATRICE Collaboration), Phys. Lett. B {\bf 478}, 31 (2000).

\bibitem{CLEOII_munu}
  M. Chada {\it et al.} (CLEO Collaboration), Phys. Rev. D {\bf 58}, 032002 (1998).

\bibitem{BaBar_munu}
  B. Aubert {\it et al.} (BaBar Collaboration), Phys. Rev. Lett. {\bf 98}, 141801 (2007).

\bibitem{L3}
  M. Acciarri {\it et al.} (L3 Collaboration), Phys. Lett. B {\bf 396}, 327 (1997).

\bibitem{opal}
  G. Abbiendi {\it et al.} (OPAL Collaboration), Phys. Lett. B {\bf 516}, 236 (2001).

\bibitem{CLEOc_taunu}
  P. Naik {\it et al.} (CLEO Collaboration), Phys. Rev. D {\bf 80}, 112004 (2009).

\bibitem{LQCD_fDs}
  A. Bazavov {\it et al.} (Fermilab Lattice and MILC Collaborations), 	 arXiv:1407.3772 [hep-lat].

\end{thebibliography}
\end{document}